\documentclass{article}
\usepackage{epsfig}
\usepackage{amsmath}
\usepackage{amssymb}
\usepackage{amsfonts}
\usepackage{array}

\begin{document}
%opening
%\begin{frontmatter}
\title{One temperature model for effective ovens}
\author{S. Tapia and J.A. del R\'{\i}o\\
Centro de Investigaci\'on en Energ\'{\i}a and \\ Centro de Ciencias de la Complejidad\\ 
Universidad Nacional Aut\'onoma de M\'exico,\\ 
Privada Xochicalco S/N, Temixco, Morelos, CP 62580, M\'exico.}

\maketitle

\begin{abstract}
Most of the thermodynamic analysis of ovens are focused on efficiency, but they need to behaves under
real-life conditions, then the effectiveness of the ovens plays a crucial role in their design.
In this paper we present a thermodynamical model able to describe the temperature evolution in ovens, furnaces 
or kilns to harden, burn or dry different products and which provides a methodology to design these heating devices.
We use the required temperature evolution for each product and process as main ingredient in the methodology and procedure to design ovens and we place in the right role the efficiency criteria. 
We use global energy balance equation for the oven under transient situation as the thermodynamic starting point for developing the model. Our approach is able to consider different configurations for these heating devices, or recirculating or open situations, etc. 
\end{abstract}
%\end{frontmatter}
\section{Introduction}
The brick kiln was a major advance in ancient technology because it provided a stronger brick than the primitive 
sun-dried product. Modern kilns or ovens are used in ceramics to fire clay and porcelain objects, in metallurgy 
for roasting iron cores, for burning lime and dolomite, and in making Portland cement between other applications.
This kind of ovens are an essential part of the manufacture of all ceramics, which, by definition, require heat treatment, 
often at high temperature. During this process, chemical and physical reactions occur which cause the material to 
be permanently altered. In the case of pottery, clay materials are shaped, dried and then fired in a oven.
Although there is an experience of some several thousands years building ovens or furnaces and we can find some high technology apparatus, very detailed numerical analysis (for example see \cite{mariasa}), two phase flow study (for instance \cite{milaneza}) or efficiency based analysis \cite{danona}; up to our knowledge there is no a simple model that allows us to calculate the physical properties and design an oven.

\begin{figure}
\begin{center}
\includegraphics[width=.8\textwidth]{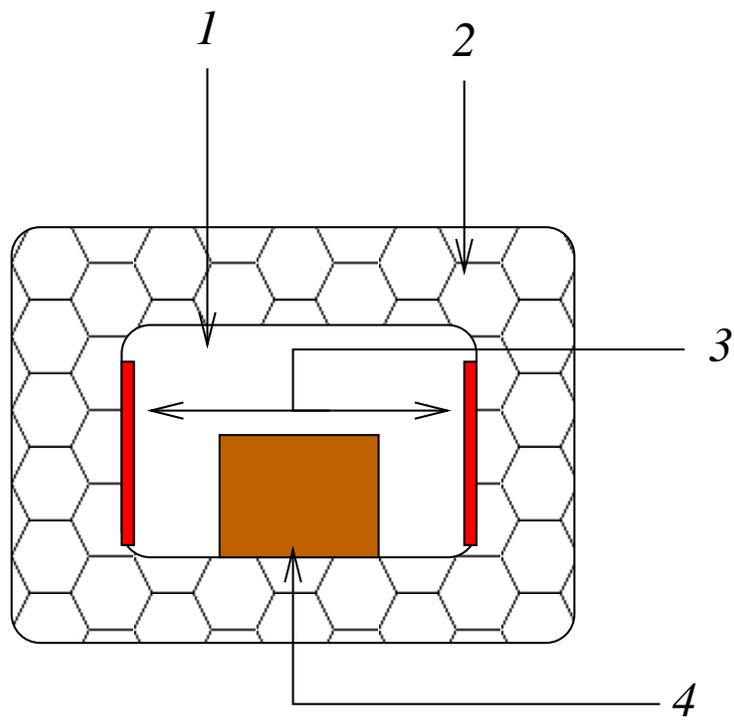}
\end{center}
\caption{Schematic oven and its main parts: 1. Chamber, 2. Thermal insulator, 3. Energy source, 4. Material to be burn.}
\label{fig1}
\end{figure}

In figure \ref{fig1} we present a scheme of our conceptual oven composed by an insulated chamber with a thermal 
energy source in which the material to be burned can be introduced. The heat can be obtained by different energy 
sources those can be inside or outside the chamber, for the purpose of this paper the energy source is not 
important. The schematic oven is composed by a cooking chamber which is thermally insulated with a high 
thermal resistivity material; inside this chamber the material to heat treated is placed. Both material and chamber are 
heated simultaneously during oven operation with internal or external energy source.

\begin{figure}
\begin{center}
\includegraphics[width=.8\textwidth]{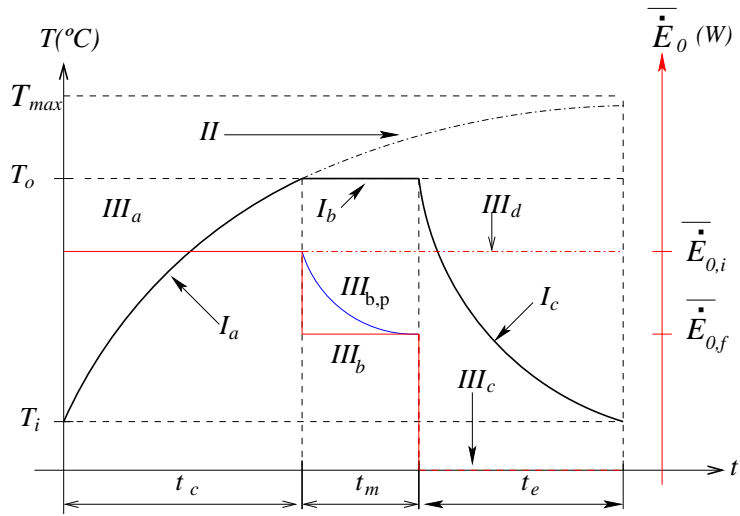}
\end{center}
\caption{Cooking process with three major phases in the heat treatment, increasing temperature, cooking period and cooling period.}
\label{fig2}
\end{figure}

The aim of this paper is not only to describe the thermal behavior of a heating device, but to 
establish a methodology for ovens design considering the temperature evolution for different products and process.

A typical temperature curve of cooking \cite{L7} is given in Fig. \ref{fig2}, where we can see that the process 
starts at $t=0$ with temperature $T_i$. If the process consists in a constant heat flow ${\bar {\dot E}}_{0,i}$ as indicated in $III_a$ 
during a time $t_c$, the internal temperature will follow a curve as $I_a$  up to reach 
$T_o$ then a maturity period follows, $t_m$ as shown in part $I_b$. Of course to follow this
plot the heat flow needs to be diminished. Paschkis and Persson~\cite{L7} proposed a slow change in heat flow as 
indicated in $III_{b,p}$. In order to obtain simplest model we will set this heat flow as constant, ${\bar {\dot 
E}}_{0,f}$ (${\bar {\dot E}}_{0,i}>{\bar {\dot E}}_{0,f}$) as plotted in curve $III_b$. After the cooking process 
 the cooling part follows with  $ {\dot E}=0$, curve $III_c$, then the temperature may be described by
 a curve like $I_c$,
at the end of cooking the thermal energy flow to the oven is canceled (curve $III_c$). With this the cooling of the oven and the products start (curve $I_c$), and the temperature ends in the temperature $T_i$ at time $t_e$. In the same figure, it is shown that if the operation temperature $T_o$ was not specified (in other words $T_o\leq T_{max}$), and the energy is introduced ${\bar {\dot E}}_{0,i}$ to the oven as shown on the lines $III_a$ and $III_d$, the thermal process starts following the temperature curve $I_a$ until time $t_c$, and later, it will continue the curve $II$. We can see that the curve $II$ has as asymptotic limit to the line of constant temperature $T_{max}$, which is reached only when $t\rightarrow\,\infty$. Of course, the graph of the figure ~\ref{fig2} has been observed in many experimental works in the firing of miscellaneous products in ovens furnaces or kilns, when the average temperature is measured, and has obtained analytically data to interpolate; however, there is not a well-founded theoretical model. 

We must emphasize the efforts to shape the curve of temperature with simple models. In 1961, Trinks and Mawhinney~\cite{L13}, modeled the temperature curve of the load when the oven is preheated and the load is introduced at room temperature. Then in 1994, Astirraga~\cite{L1} presented a simple model for the thermal development of an electric oven, and he founded the differential equation that describes the process of warming, however, the initial conditions of the thermal process were not well established. Therefore, the found solution  does not correspond to the baking process, and this only describe qualitatively the process. In 2002, Abraham and Sparrow~\cite{L0}, described the thermal behavior of a load at room temperature introduced on an electric preheated oven; the analytic relation found has similarity with Trinks and Mawhinney's work~\cite{L13}. In 2005, Tapia and del R\'io~\cite{A14} characterized and model theoretically a ''solar cooker'' finding that the analytic representation of the temperature evolution inside the cooking chamber of the cooker depends on the thermal resistance of the oven, its heat capacity, and the characteristic parameters of the process. They also showed that cooking of food had a thermal development according to the predictions of theoretical models. In 2007 Schwarzer and Vieira~\cite{A12} developed a theoretical model, to evaluate the thermal development in a pot heated by solar energy, however, they apply the relation for two different cases, the first one for cooking in a pot exposed to the atmosphere, and that receives the solar flux concentrated by the reflect of a mirror, and the second one to a solar oven box. We should emphasize that, according to their approach, their theoretical model, can only describe the balance of energy in the pot exposed to the environment, and not to the pot inside the oven. 

The goals of this paper are: develop a methodology to design ovens based on a simple thermodynamic model that describes the temperature evolution according with the heating process indicated in figure.~\ref{fig2}.

\section{Modeling typical heating curves in ovens}
In this section we present a theoretical model for describing the temperature evolution inside of an oven in terms of some parameters that allow to us to design ovens or furnaces according to specific cooking requirements.

\subsection{Describing temperature evolution.}
We wish to describe the evolution of a representative temperature inside the cooking chamber of a oven. 
Here it is important to mention that a detailed description of the temperature field may be important for some specific processes or for big cooking chambers where stratified temperature conditions could be found.  However
in this first approach, we can assume that we are selecting a good place to measure the temperature and it is representative of the whole temperature field, with this representative temperature we can construct an one temperature model for describing the cooking process as we have done above.
In order to do the one temperature model we will use a global balance of energy using the cooking chamber as the control volume. Due to the fact that, we try to reproduce the cooking curve, shown in figure~\ref{fig2}, we will consider the next statements: 
\begin{enumerate}
\item We consider that all objects in the chamber are under thermal equilibrium condition. 
\item The materials that compose the oven and the cooking material have the constant physical properties, i.e. they do not depend on temperature. 
\item The heat capacity $C_{j}=c_{p,j}m_j \omega_j$ of insulators and supports of the oven are considered in the balance energy equation, where $c_{p,j} $, $m_j$ are the specific heat and the mass of the component $j$ in the oven.
\item The change of temperature $T$ with respect to $t$ has the following requirements: $\frac{dT}{dt}$ is maximum at $t=0$, it is a non increasing function of $t$ and $\frac{dT}{dt} \rightarrow\, 0$ when $t\rightarrow\,\infty$.
\item We neglect the energy of the chemical or phase transformations.
\item We assume that no work enters or goes out from the furnace
\end{enumerate}
\begin{enumerate}
\item We consider that cavity and both air and body surfaces places in it are in thermal equilibrium.
\item The physical properties, as thermal conductivity and specific heat of oven components and cooking materials, are constant. 
\item The temperature change in the inner part of the oven satisfies the following restrictions:
\begin{itemize}
\item $ \frac{dT}{dt}= \varepsilon\geq0$.
\item $\varepsilon$ is small in the heating process.
\item $\varepsilon(t)$ is maximum at $t=0$, i.e., it is a decreasing function on $t$.
\item $\varepsilon (t)\rightarrow\, 0$ when $t\rightarrow\,\infty$.
\end{itemize}
\item We consider the following approximations for the cooling part.
\begin{itemize}
\item $ \frac{dT}{dt}= \varepsilon\leq0$.
\item $\mid\varepsilon\mid$ is small at $t=t_c+t_m$.
\item $\mid\varepsilon(t)\mid$ is maximum at $t=t_c+t_m,$ i.e., it is a decreasing function on $t$.
\item $\mid\varepsilon (t)\mid\rightarrow\, 0$ when $t\rightarrow\,\infty$.
\end{itemize}
\item We are neglecting all the energy involved in chemical changes
\end{enumerate} 

Following these ideas, the energy balance equation of the oven can be expressed as:~\cite{L10}-\cite{L12}:

\begin{equation}
{\dot Q}_{e}-{\dot Q}_{s}=\dfrac{d E_{ac}}{dt},
\label{eq:1}
\end{equation}
where ${\dot Q}_e$ is the input heat, ${\dot Q}_s$ is the heat diffused into the environment, $E_{ac}$ is the accumulated energy of the furnace.

Here we can use that the accumulated energy can be expressed as
\begin{equation}
 \dfrac{d E_{ac}}{dt}=C_{T}\frac{dT}{dt}
\label{eq:2}
\end{equation} 
where $C_{T}= \sum_{j}c_{p,j}m_j\omega_j$ is the effective heat capacity of the oven. Here we are considering the supports, insulator materials, meals, pastes and any other component as $j$ elements.

In this work we propose a simple model to follow a cooking procedures according with Ia and II curves in figure \ref{fig2}. This model will help us to design an effective furnace for product cooking with a constant energy flow input 
implemented by a control system that will have the characteristic of adjusting the energy flow input to a constant value ${\dot Q}_{e}={\bar{\dot E}}_{0}$. Here it is important to mention that 
$ {\bar{\dot E}}_{0}$ includes all energies we are supply to the furnaces, i.e. ${\bar{\dot E}}_{0}$ involves 
electrical power in case of electrical ovens, or heat coming from combustion in case of gas ovens, or solar power in case of solar furnaces or any combination of the former for hybrid energy devices. With this assumption we can write
\begin{equation}
{\bar{\dot E}}_{0}-{\dot Q}_{s}=C_{T}\frac{dT}{dt},
\label{eq:5}
\end{equation} 
from Eq. (\ref{eq:1}) to describe the temperature of the oven. 

Within this approximation, and according to statements 1 to 5, we will estimate the maximum value for ${\dot Q}_s$ by the relation that evaluates heat flow in steady state, which can be expressed mathematically as:
\begin{equation}
{\dot Q}_s=\frac{T-T_a}{R_T},
\label{eq:4}
\end{equation}  
where $T_a$ is the environment temperature, and $R_T$ is the total thermal resistance from the chamber to the environment. Substitution of Eq. ~(\ref{eq:4}) in Eq. (\ref{eq:5}) gives
\begin{equation}
{\tau} \frac{dT}{dt}+T=T_{max},
\label{eq:13}
\end{equation}
where we have defined:
\begin{equation}
\tau =R_{T}C_{T}.
\label{eq:14}
\end{equation}
and 
\begin{equation}
T_{max}={\bar{\dot E}}_{0}R_{T}+T_{a},
\label{eq:15}
\end{equation} 
here $\tau$ and $T_{max}$ are constant.
The solution of this equation is
\begin{equation}
T(t)=T_{max}\left( 1-\exp\left[ \frac{-t}{\tau}\right] \right) +T_{a}\exp\left[ \frac{-t}{\tau}\right] .
\label{eq:18}
\end{equation}
The relation~(\ref{eq:18}) describes the thermal behavior within the chamber, and by selection specific parameters $\tau$ and $T_{max}$ it displays a similar curve to the one shown in the figure~\ref{fig2} with the curves $Ia$ y $II$. 

To be able to observe the behavior of the difference temperatures between the maximum temperatures $T_{max}$ and the room temperature $T_a$, we substitute~(\ref{eq:15}) in the relation~(\ref{eq:18}), and after doing algebraic simplifications we find this relation:
\begin{equation}
T(t)-T_{a}={\bar{\dot E}_0}R_T\left( 1-Exp\left[ \frac{-t}{\tau}\right] \right),
\label{eq:DIF}
\end{equation}
If we use $\Theta=T-T_a$ and $\Theta_{max}={\bar{\dot E}_0}R_T$, then relation~(\ref{eq:DIF}) can be written as
\begin{equation}
\Theta(t)=\Theta_{max}(t)\left( 1-Exp\left[ \frac{-t}{\tau}\right] \right) .
\label{eq:DIF2}
\end{equation} 

\begin{figure}
\begin{center}
\includegraphics[width=.8\textwidth]{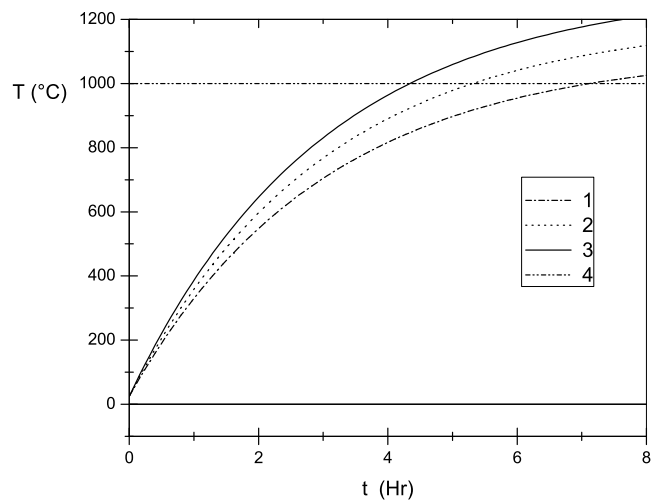}
\end{center}
\caption{Temperature evolution with $\tau=3\,Hr.,$ and different $T_{max}$ ($^0 C$)values, line 1, $1100$; line 2, $1200$; line 3,  $1300$. }
\label{fig:3}

\end{figure}

As an example we graph the equation~(\ref{eq:18}) varying $\tau$ and considering $T_{max}$ constant, and viceversa, we have also considered that the furnace cooks products at a temperature of $T_o=1000^{\circ}C$, and the results are shown in the figures \ref{fig:3} and 4, which when compared with the normal thermal performance of the curves $I_a$ and $II$ in the figure~\ref{fig1} show good agreement. 

\begin{figure}
\begin{center}
\includegraphics[width=.8\textwidth]{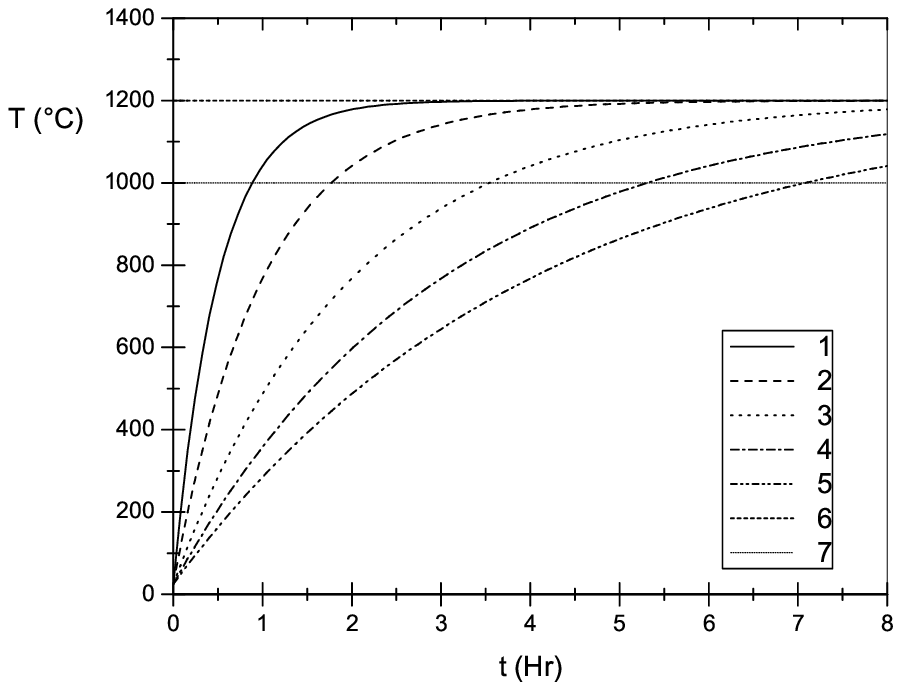}
\caption{Theoretical performance of the furnace, considering constant $T_{max}=1200^{\circ}C$, and varying $\tau$. The lines: 1. $\tau=0.5 hr$; 2. $\tau=1 hr$; 3. $\tau=2 hr$; 4. $\tau= 3 hr$; 4. $\tau=4 hr$.}
\end{center}
\label{fig4a}

\end{figure}

To model the maturation curve $I_b$ is required to supply a constant energy ${\bar{\dot E}}_{0,f}$ , that compensates the energy dissipated to the environment. 
For this reason to obtain $T(t)=T_o$ during the time $t$ in the interval $t_c\,\leq t\,\leq \,t_c+t_m$, is required:
\begin{equation}
{\bar{\dot E}}_{0,f}={\frac{T_o-T_a}{R_T}}.
\label{eq:EnerDis}
\end{equation} 
Finally to model the cooling curve $I_c$ in the interval $t_c+t_m \leq t \leq t_c+t_m+t_e$, we must consider ${\bar{\dot E}}_{0}=0$ in the balance equation~(\ref{eq:5}), in which, we must also substitute the relation~(\ref{eq:4}), therefore we obtain: 
\begin{equation}
-{\frac{T-T_a}{R_T}}=C_T{\frac{dT}{dt}}, 
\label{eq:BalEnf}
\end{equation}
the relation~(\ref{eq:BalEnf}) has the solution:
\begin{equation}
T=T_a+\left( T_o-T_a \right)Exp\left(\frac{t_c+t_m}{\tau}\right) Exp\left( \frac{-t}{\tau}\right),
\label{eq:SolEnf1}  
\end{equation}
or the relation:
\begin{equation}
\Theta=\Theta_{o} Exp\left(\frac{t_c+t_m}{\tau}\right)Exp\left( \frac{-t}{\tau}\right),
\label{eq:SolEnf2}  
\end{equation}
where $\Theta_{o}=T_o-T_a$. 

Here we need to emphasize that ${\bar{\dot E}}_{0,i} $ can be provided to the furnace using different energy sources, it can be supplied by electricity, or heat from combustion, or radiation, or any combination of them and some other energy source.

Once we have modeled the cooking curves, $I_a$, $I_b$ and $I_c$; and also $I_a$ along with $II$, and defining the way of joining the different energy sources, we will use the development of furnace sizing.

Now we calculate the change of energy accumulated in the furnace through the relation:
\begin{eqnarray}
\Delta{E}_{ac}=C_T*\Theta(t)=C_T*{\bar{\dot E}_0}R_T\left( 1-Exp\left[\frac{-t}{\tau} \right] \right)=\nonumber\\=\tau{\bar{\dot E}}_0 \left( 1-Exp\left[\frac{-t}{\tau} \right] \right),
\label{eq:DeltaAcum}
\end{eqnarray} 
The total energy that enters the furnace according to the model is:
\begin{equation}
E_T={\bar{\dot E}}_0 t,
\end{equation} 
therefore the energy dissipate to the room is:
\begin{eqnarray}
Q_s=E_T-E_{ac}={\bar{\dot E}}_0 \left( t-\tau\left( 1-Exp\left[\frac{-t}{\tau} \right] \right)\right).
\label{eq:DIS}
\end{eqnarray}
We can observe that according to these results, the energy dissipate to the room is an increasing function on time, and the accumulated energy is bounded, which is in perfect compatibility with the physical processes of the furnaces operation.

\subsection{Furnace design.}
We use the previous ideas to design a furnace, according to the figure~\ref{fig2}, with a rectangular form.
we must know,
\begin{itemize}
\item The maximum operating furnace temperature $T_{o}$.
\item Cooking time $t_c$.
\item Maturing time $t_m$.
\item The temperature $T_a$.
\end{itemize}
With this information, we determine the furnace parameters: thickness of insulator $\delta$, ${\bar{\dot E}}_{0,i}$, ${\bar{\dot E}}_{0,f}$, $\Theta_{max}$, $R_T$, $C_T$, $\tau$ and $T_{max}$ to can describe the thermal behavior of the oven.

The parameter $\tau$ determines the speed of temperature evolution, and while heating or cooking products we can observe two different types of thermal evolution, as it can be seen in the figure~\ref{fig4a}, a slow temperature curve where $\tau$ has a value in the interval:
\begin{equation}
\frac{2}{5} t_c <\tau\leq \frac{4}{5} t_c,
\label{eq:tauTm}
\end{equation} 
and a fast temperature curve when $\tau$ has a value on the interval: 
\begin{equation}
\frac{1}{10}\,t_c\,<\tau\leq \frac{2}{5}\,t_c.
\label{eq:tauTm2}
\end{equation}

$R_T$ and $C_T$  are the parameters that control the thermal behavior, for this, it must be adjusted to fit the relation:
\begin{equation}
C_T=\frac{\tau}{R_T}=\frac{\alpha t_c}{R_T},
\label{eq:CT2}
\end{equation} 
where $\alpha$ is the selected in the interval $\frac{1}{10}<\alpha<\frac{4}{5}$ and $t_c$ from relations (\ref{eq:tauTm}) and~(\ref{eq:tauTm2}). We must also take into account the relations~(\ref{eq:2}) and ~(\ref{eq:CT2}) to fulfill the requirement:
\begin{equation}
\sum_j c_{p,j}m_j\omega_j=\sum_j c_{p,j}\rho_jV_j\omega_j=\frac{\alpha t_c}{R_T},
\label{eq:AjusC}
\end{equation}
where the material $j$ has the density $\rho_j$, mass $m_j$ and the volume $V_j$. We will define $j$ materials and their properties according to the table~\ref{fig:CT}. 
\begin{table}
\begin{center}
% use packages: array
\begin{tabular}{|c|c|c|c|c|c|c|}
\hline Index $j$ & Material & $\rho\,(kg/m^{3})$ & $C_{p,j}\,(J/kg^{\circ}C)$ & $V_j\,(m^{3})$ &$m_j$& $w_j$  \\ \hline
\hline$1$ & Cooking material & $\rho_{1}$ & $c_{p,1}$ & $V_1$ &$m_1$& $w_1=1$\\ 
\hline$2$ & Heat-resistant & $\rho_{2}$ & $c_{p,2}$ &$V_2$ &$m_2$& $w_2=1$\\ 
\hline$3$ & Thermal insulator & $\rho_{3}$ & $c_{p,3}$ & $V_3$&$m_3$ & $w_3=0.55$ \\ 
\hline
\end{tabular}
\end{center}
\caption{Defined parameter for the evaluation of $C_T$ \cite{L1}. }
\label{fig:CT}
\end{table}

Also we can determine $\Theta_{max}$ by considering the value of $\tau_0=\alpha t_c$ and that the cooking temperature $\Theta_o$ is reached at the time $t=t_c$, after, when replacing in the relation~(\ref{eq:DIF2}), we get:
\begin{equation}
\Theta_o=\Theta_{max}\left[ 1-Exp\left( \frac{-1}{\alpha}\right) \right],
\label{eq:ThetaMax} 
\end{equation} 
where $\Theta_o\,=\,T_o-T_a$. from the relation~(\ref{eq:ThetaMax}) when doing some algebra we get to:
\begin{equation}
\Theta_{max}=\frac{\Theta_o}{\left[ 1-Exp\left( \frac{-1}{\alpha}\right) \right]},
\label{eq:ThetaMax2} 
\end{equation} 

It is also important to determine $T_{max}=\Theta_{max}+T_a$ 
and if we replace the value of $\Theta_{max}$ from the relation ~(\ref{eq:ThetaMax2}) we obtain:
\begin{equation}
T_{max}=T_a+\frac{\Theta_o}{\left[ 1-Exp\left( \frac{-1}{\alpha}\right) \right]}.
\label{eq:TMax} 
\end{equation}  

To determine the energy value for the curve $I_a$ we know that 
$${\bar{\dot E}}_0\,={\bar{\dot E}}_{0,i}\,=\,\frac{\Theta_{max}}{R_T}.$$ 
This relation is the adequate energy flow for the furnace operation at a constant energy, and we must equal it to the relation of the energy flow of the sources, 

The energy ${\bar{\dot E}}_{0,f}$ needed to preserve the temperature during the time interval  $t_c\, \leq t\, \leq \,t_c+t_m$ is evaluated while considering the relation~(\ref{eq:ResTot}) and the cooking conditions, therefore we obtain:
\begin{equation}
{\bar{\dot E}}_{0,f}=\frac{( T_{o}-T_a) }{R_T}.
\label{eq:EnerMad}
\end{equation}

Finally, the temperature $T(t)$ of the furnace, in the interval $0\leq\,t\leq\,t_c$, according to the model will have the relation:
\begin{equation}
T(t)=\left(T_a+\frac{\Theta_o}{\left[ 1-Exp\left( \frac{-1}{\alpha}\right) \right]} \right) \left( 1-Exp\left[ \frac{-t}{\alpha t_c}\right]  +T_{a}Exp\left[ \frac{-t}{\alpha t_c}\right]\right),
\label{eq:18B}
\end{equation} 
to evaluate $\Theta(t)$ we can do it through the relation:
\begin{equation}
\Theta(t)=\left( \frac{\Theta_o}{\left[ 1-Exp\left( \frac{-1}{\alpha}\right) \right]}\right) \left( 1-Exp\left[ \frac{-t}{\alpha t_c}\right] \right).
\label{eq:DIF3}
\end{equation}

In the interval $t_c\leq\,t\leq\,t_c+t_m$, will have the constant temperature $T(t)=T_o$. In this interval, the cooling temperature will evolve following the relation:
\begin{equation}
T(t)=T_a+\left( T_o-T_a\right)Exp\left(\frac{t_c+t_m}{\tau_0}\right)Exp\left( \frac{-t}{\alpha\,t_c}\right),
\label{eq:TemHOR}  
\end{equation}  
or because of the temperature $\Theta$ like:
\begin{equation}
\Theta(t)=\Theta_o\,Exp\left(\frac{t_c+t_m}{\tau_0}\right)Exp\left( \frac{-t}{\alpha\,t_c}\right).
\label{eq:TemHOR2}  
\end{equation} 

Once all the parameters required to characterize a square furnace have been met, we will present conclusions.

\section{Oven sizing}

Here we give some examples using different geometries as rectangular, cylindrical and semi-spherical. 

\begin{figure}
	\centering
	\includegraphics[width=.8\textwidth]{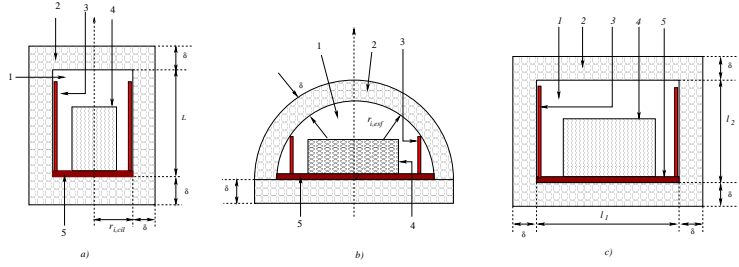}
	\caption{Standard oven shapes: a). cylinder., b) semi-spherical, c) rectangular. Cooking chamber 1) thermal insulator 2) energy source 3) cooking material 4)dead mass 5) refractory support. }
	\label{fig:Hornos-formas-Reg}
\end{figure}

We have developed the following design steps to determine all the parameters involved in the oven
\begin{enumerate}
\item Let determine $t_c$ and $T_o$ according with the typical cooking curve.
\item Let $T_i$ the initial temperature, taken as $T_{a}$ the ambient temperature.
\item From  $T(t)$ curve describing the expected process we can select $\tau=\tau_0$. 
\item The geometrical form of the oven is selected according with practical information of the specific process. 
\item Formulas for $R_T(\delta)$ and $C_T(\delta)$ are determine according with the geometrical form and $\delta=\delta_0$ is a fitting parameter to be determined from
\begin{equation}
\tau_0=C_T(\delta)R_T(\delta).
\label{eq:TaUU}
\end{equation}
\item With $\delta_0$ we can obtain $R_T(\delta_0)$ and $C_T(\delta_0)$.
\item With these data we can compare  $T_{max}$, $\overline{\dot E}_0$, $\overline{\dot E}_{0,m}$, etc.  and the curve $T(t)$ with the expected curve.
\item After some iterations we can get the final geometrical parameters for the oven.
\end{enumerate}

Thus, in order to sizing the oven we need to consider relations \ref{eq:CT2} and \ref{eq:AjusC} to determine the maximum load. From these relations we observe that $\tau (\delta)$, then to determine the adequate thickness we need to propose that

$$
\tau_o=\tau (\delta_0) = \alpha t_c. 
$$
where $\delta_o$ is the selected thickness of the insulator such as with this value the corresponding $ \tau_0$ to follow the cooking process we select.
 
$$
R_T=\frac{R_{e,p}}{\bar{A}_p},
$$
Where $\bar{A}_p$ is the average area of the wall and $ R_{e,p}$ is the thermal resistance per area unit. In the case of a rectangular oven with lengths $l_1,l_2,l_3$ and thickness $\delta$ we can get:
\begin{equation}
R_T=\frac{\frac{\delta}{k_{ais}}+\frac{1}{h}}{2\left(l_1 l_2 + l_2 l_3 + l_1 l_3\right) + 4\left( l_1+l_2 +l_3\right)\delta +  12 \delta^2 }.
\label{eq:ResTot}
\end{equation} 
This expression for the case of a cube is transformed in

\begin{equation}
R_T=\frac{\frac{\delta}{k_{ais}}+\frac{1}{h}}{6l^2+12l\delta+12\delta^2}.
\end{equation} 

In the case of a cylinder close by to parallel walls, we can write
\begin{equation}
R_{T,cil}=\frac{1}{\frac{\bar A_{p,cil}}{R_{e,p}}+\frac{{\bar L}}{R_{e,cil}}}=\frac{1}{\frac{\pi\left[ 2\,r_{i,c}\left( r_{i,c}+\delta\right)+ \delta^{2}\right] }{R_{e,p}}+\frac{L+\delta}{R_{e,cil}}},
\label{eq:ResTotcil}
\end{equation}  
and in this case the specific $R_{e,cil}$ is
\begin{eqnarray}
R_{e,cil}=\frac{Ln(\frac{r_{i,cil}+\delta}{r_{i,cil}}) }{2\,\pi\,k_{ais}}+\frac{1}{2\,\pi\,\left( r_{i,cil}+\delta\right) \,\,h\,},
\end{eqnarray} 
where $r_{i,cil}$ is the internal an external radius of the cylinder and $r_{i,cil}$ is the internal radius of the insulator.

In the case of a semi-spherical oven we have that
\begin{equation}
R_{T,esf}=\frac{1}{\frac{{\bar A}_{p,esf}}{R_{e,p}}+\frac{{1}}{R_{e,esf}}}=\frac{1}{\frac{\pi\left( 2\,r_{i,esf}\left( r_{i,esf}+\delta\right)+\delta^{2} \right) }{2R_{e,p}}+\frac{{1}}{R_{e,esf}}},
\label{eq:ResTotesf}
\end{equation} 
where $r_{i,esf}$ is the radius of the sphere and the thermal resistance can be obtained as
\begin{equation}
R_{e,esf}=\frac{\delta}{4\,\pi\,\left( r_{i,cil}+\delta\right) \,r_{i,cil}\,k_{ais}}+\frac{1}{4\,\pi\,\left( r_{i,esf}+\delta\right)^{2} \,h},
\end{equation}
where $r_{i,esf}$ is the internal radius of the sphere.

As we need the heat capacities of the insulator for different shapes, in these cases we have:
\begin{enumerate}
 \item Parallelepiped shape.

Considering $l_1$, $l_2$ and $l_3$ as the dimensions of the parallepiped and $\delta$ as the thickens of the insulators we get 
\begin{equation}
C_{T,p}=C_{x,p}+c_{p,ais}\rho_{ais}g_{ais}\left[8\delta^{3}+4(l_1+l_2+l_3)\delta^{2}+2(l_1l_2+l_2l_3+l_1l_3)\delta \right] 
\label{eq:CTotPLL}
\end{equation} 
where $c_{p,ais}$, $\rho_{ais}$ and  $\delta$ are the mean heat capacity, density and thickness of the insulator, respectively. Here $C_{x,p}$ is heat capacity of the parallelepide oven without considering the insulator term.
\item Cubic shape.
Considering $l$ as the size length, then, 
\begin{equation}
C_{T,cub}=C_{x,cub}+c_{p,ais}\rho_{ais}g_{ais}\left[8\delta^{3}+12l\delta^{2}+6l^{2}\delta \right] 
\label{eq:CTotCub}
\end{equation} 
Here $C_{x,cub}$ is heat capacity of the cubical oven without considering the insulator term.
\item Cylinder shape.
Here $L$ and $r_{i,cil}$ are the length and radius of the cylinder, then, 
\begin{equation}
C_{T,c}=C_{x,c}+c_{p,ais}\rho_{ais}g_{ais}\left[\pi\left(2\delta^{3}+(L+4r_{i,cil})\delta^{2}+2r_{i,cil}(L+1)\delta \right)\right] 
\label{eq:CTotCil}
\end{equation} 
where again $C_{x,c}$ is heat capacity of the cylindrical oven without considering the insulator term.
\item Semi-sphere.
With $r_{i,esf}$ as radius 
\begin{equation}
C_{T,e}=C_{x,e}+c_{p,ais}\rho_{ais}g_{ais}\left[\pi\left(\frac{7}{3}\delta^{3}+6r_{i,esf}\delta^{2}+5(r_{i,esf})^{2}\delta \right)\right] 
\label{eq:CTotEsf}
\end{equation} 
where again $C_{x,e}$ is heat capacity of the semi-spherical oven without considering the insulator term.
\end{enumerate}
With these formulas our description of the time evolution temperature inside the oven is complete.

\subsection{Example of a electrical oven}
\begin{figure}
	\centering
	\includegraphics[width=.9\textwidth]{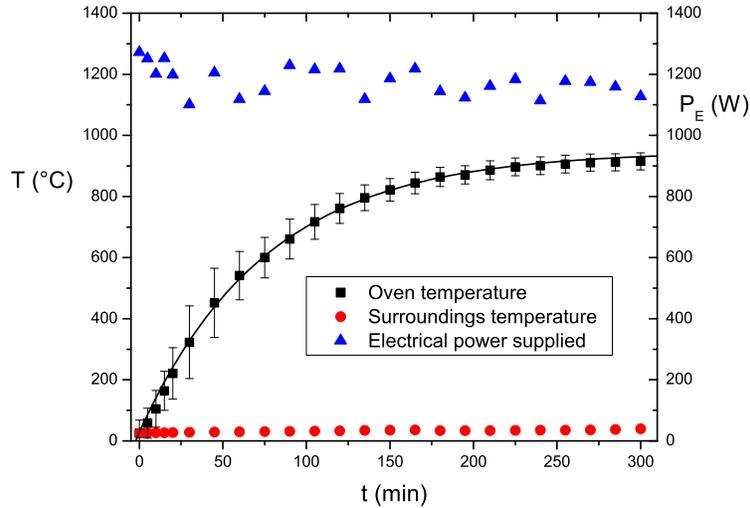}
	\caption{Comparison between experimental data and themodynamical model}
	\label{fig:Grafica-Temperatura-Horno}
\end{figure}

Using this methodology we have designed an electrical furnace. The corresponding parameters are given in Table \ref{fig:Capacidades-Teo-Finales}. This furnace was use to cook a ceramic piece. During the cooking process the temperature was taken and they are plotted  in fig.~\ref{fig:Grafica-Temperatura-Horno}, as it can be seen, the agreement between model and real data is good.

\begin{table}
\begin{center}
\begin{tabular}{||c|c|c|c|c|c|c||}
\hline Component & $\rho_j\,(\frac{kg}{m^{3}})$ &$v_j$ $(m^{3})$ & $m_j\,(kg)$ & ${\overline c}_{p,j}(J/kg \cdot K)$ & $w_j$ &$C_{j}(J/K)$ \\ 
\hline \hline Product & $\rho_1$  & $v_1$ & $0.380$ & $820$ & $1$ & $311$\\ 
\hline Heat-resistant & $\rho_{2}$  & $v_{2}$ &$2.650$ & $820$ & $1$ & $2.146$\\
\hline Electrical resistance & $\rho_3$  & $v_3$ & $0.3$ & $450$ & $1$ & $1.50$\\
\hline Structure & $\rho_5$  & $v_5$ &$8.0$ & $450$ & $0.05$ &$180$ \\
\hline Insulator & $\rho_6$  & $v_6$ &  $6.4$ & $1000$ & $0.45$ &$2880$ \\  
\hline \hline
\end{tabular}
\end{center}
\caption{Physical properties of the oven\cite{L1} .}
\label{fig:Capacidades-Teo-Finales} 
\end{table}

\section{Conclusions}
In this paper we present a simple model to describe the mean  temperature inside an oven considering global thermal balances. With this model we analyze the thermal behavior under changes of the physical properties of the oven components. 

Since different products require different temperature time evolution in the cooking process, we present a methodology to design effective ovens, i.e., ovens, that from a global point of view, cook following a specific temperature evolution. Here is important that the criterion is in the effectiveness instead the efficiency, because it is more important following the required temperature than the use of the energy in the cooking process.

We have considered different shapes and present examples of our methodology. 

We test this methodology performing an experiment under specific parameters and comparison between theoretical and experimental results is good.

\textbf{Acknowledgements}
This work was partially supported by DAGAPA-UNAM under project IN106210.

\end{document}